\documentstyle[12pt]{article}
\def\be{\begin{equation}}
\def\ee{\end{equation}}
\def\bea{\begin{eqnarray}}
\def\eea{\end{eqnarray}}
\advance\textheight by \topskip
\begin{document}
\begin{center}
{\LARGE \bf Class of Einstein--Maxwell Dilatons}
\vskip1cm
{\bf Tonatiuh Matos$^{\ast}$, Dar\'\i o N\'u\~nez$^{\dagger}$
and Hernando Quevedo$^{\dagger}$}\\
\vskip5mm
$^{\ast}$  Departamento de F\'\i sica,
Centro de Investigaci\'on y Estudios Avanzados del I.P.N.\\
A.P. 14-700, \  07000 M\'exico D.F.\\
\vskip3mm
$^{\dagger}$ Instituto de Ciencias Nucleares, UNAM, \\Circuito Exterior CU,
A.P. 70-543, M\'exico, D. F. 04510,  M\'exico. \\
\end{center}

\begin{abstract}
{Three different classes of static solutions of the Einstein--Maxwell
equations non--minimally coupled to a dilaton field are presented.
The solutions are given in general in terms of two
arbitrary harmonic functions and involve among others
an arbitrary parameter which determines their applicability as
charged black holes, dilaton black holes or strings. Most of the
known solutions are contained as special cases and can be non--trivially
generalized in different ways.}
\end{abstract}

\vfill\eject
\section{Introduction}

The study of the gravitational interaction coupled to the Maxwell and
dilaton fields has been the subject of recent investigations.
Einstein--Maxwell fields of black hole type are probably the most
interesting objects predicted by classical general relativity.
Dilaton fields appear (coupled to Einstein--Maxwell fields) naturally in the
low-energy limit of string theory and as a result of a dimensional
reduction of the Kaluza--Klein Lagrangian. Therefore, the study of the
dilaton field coupled to matter is of importance for the understanding
of more general theories.
In this work we investigate the Lagrangian density \cite{HH1}
\be
{\cal L} = \sqrt{-g} \left [ - R + 2 (\Delta \Phi)^2 +
e^{-2\alpha \Phi} F^2 \right] \ , \label{lag}
\ee
where $g=\det(g_{\mu\nu}), \ \mu, \nu = 0,1,2,3$, $R$ is the
scalar curvature, $F_{\mu\nu}$ is the Maxwell field, and
$\Phi$ is the dilaton.  The constant $\alpha\geq 0$ determines the
special theories contained in Eq.~(\ref{lag}).
For $\alpha= \sqrt{3} $, the Lagrangian
(\ref{lag}) leads to the Kaluza--Klein field equations obtained from the
dimensional reduction of the five--dimensional Einstein vacuum equations.
For $\alpha = 1$, the Lagrangian concides with the low energy limit of string
theory with vanishing dilaton potential \cite{HH2}.  Finally,
in the extreme limit $\alpha=0$, Eq.~(\ref{lag}) reduces to
the Einstein--Maxwell
theory minimally coupled to the scalar field.

The field equations obtained from Eq.~(\ref{lag}) are
\bea
(e^{-2\alpha\Phi} F^{\mu\nu})_{;\mu} &=& 0 \label{fe1} ,\\
\Phi^{;\mu}_{\ ;\mu}
+ {\alpha\over 2} e^{-2\alpha\Phi} F_{\mu\nu}F^{\mu\nu} &=& 0\ , \label{fe2}\\
R_{\mu\nu} = 2 \Phi_{,\mu} \Phi_{,\nu} &+& 2 e^{-2\alpha \Phi} (F_{\mu\lambda}
F_\nu^{\ \lambda} - {1\over 4} g_{\mu\nu}
F_{\alpha\beta}F^{\alpha\beta} ) \ , \label{fe3}
\eea
where a semicolon denotes the covariant derivative with respect to
$g_{\mu\nu}$ and a comma represents partial differentiation. A few exact
solutions of Eqs.(\ref{fe1}-- \ref{fe3}) are known which reveals
many interesting features
of the dilaton field (see ref. \cite{HH1} and the literature cited therein).
In this paper, we obtain several solutions to Eqs.(\ref{fe1}--\ref{fe3})
by applying
the potential space formalism originally developed by Neugebauer and
Kramer for Einstein--Maxwell fields \cite{NK}.

We assume that the spacetime is characterized by two Killing vector
fields, $X$ and $Y$, and introduce  coordinates $t$ and $\phi$ which
are chosen as $X=\partial/\partial t$ and $Y=\partial/\partial \phi$.
The corresponding line element can then be expressed as
\be
ds^2 = f(dt - \omega d\phi)^2 - f^{-1}[ e^{2 k} (d\rho^2 + dz^2) + \rho^2
d\phi^2] \ , \label{eq:lel}
\ee
where $f, \ \omega$ and $k$  are functions of $\rho$ and $z$ only.

In section 2 we introduce the abstract potential space that leads to a set
of equations equivalent to the field equations (\ref{fe1}--\ref{fe3}) for
the metric (\ref{eq:lel}).
It turns out that for special values of the parameter $\alpha$, the main
field equations may be reduced to
a chiral equation  which can be solved by using the method of harmonic maps.
In section 3 we present two classes of exact solutions the first of which
describes a general static gravitational field coupled to a dilaton field,
both of them being determined by an arbitrary harmonic function. The second
class of solutions contains two harmonic functions which determine the
gravitational, dilaton and electric fields. Several particular solutions
of these classes are derived and briefly discussed.

\section{The Potential Space}

Due to the non--minimal coupling of the electromagnetic and dilaton
fields in the Lagrangian (\ref{lag}), the explicit form of the field equations
for the line element (\ref{eq:lel}) becomes very cumbersome. It turns
out that
the metric functions $f$ and $\omega$ must satisfy two partial, second
order differential equations which depend also on the electromagnetic
and dilaton fields. Moreover, the function $k$ satisfies two partial,
first order differential which, however, can be integrated by quadratures
once $f$, and $\omega$ are known. In general, it is very hard to obtain
directly exact solutions to this system. Therefore, we will apply here a
simplifying
approach which is based upon the introduction of an abstract space with
coordinates defined by the metric functions entering (\ref{eq:lel}).
To obtain
the ``metric" that determines the abstract space, we introduce the line
element (\ref{eq:lel}) into the Lagrangian (\ref{lag}) and neglect the total
divergence terms
which contain second order derivatives. The resulting Lagrangian depends on
the metric functions as well as their first order derivatives, and may be
handle like a Lagrangian for a mechanical system with coordinates which
coincide with the metric functions of (\ref{eq:lel}). We then apply a Legendre
transformation involving all the cyclic coordinates of the Lagrangian.
This procedure is similar to that used in classical mechanics to obtain
the Routh function, and has recently been used in other cases \cite{NQ1},
\cite{NQ2}.
The resulting Lagrangian is then the metric determining the
abstract space, and may be written as
\be
{\cal L} = dS^2 = {\rho\over 2f^2}[ Df^2 + (D\epsilon + \psi D\chi)^2]
+ {\rho \kappa^2\over 2f}\left( D\psi^2 + {1\over \kappa^4} D\chi^2 \right)
+ {2\rho \over \alpha^2\kappa^2 } D\kappa^2 \ , \label{eq:lag1}
\ee
where $D=(\partial/\partial\rho,\  \partial/\partial z)$ is a vector operator
with its ``dual" $\tilde D = (\partial/\partial z, \ -\partial/\partial\rho)$
such that $D\tilde D G(\rho,z)= 0$ for any arbitrary function $G(\rho,z)$.
The ``coordinates" of (\ref{eq:lag1}) are defined by the following equations
\be
\psi = 2 A_t \ , \quad \kappa^2 = e^{-2\alpha\Phi}\ , \quad
\tilde D\chi = 2{f\kappa^2 \over \rho}(\omega DA_t + DA_\phi)\ , \quad
\tilde D\epsilon  = {f^2\over \rho} D\omega + \psi \tilde D\chi \ ,
\ee
where the electromagnetic vector potential $A_\mu = (A_t, 0 , 0 , A_\phi)$
has two non--vanishing components only, in accordance to the symmetry
properties of the gravitational field.  According to their definition, the
coordinates $f, \epsilon, \psi, \chi$ and $\kappa$ entering
Eq.~(\ref{eq:lag1})
may also
be interpreted as the gravitational, rotational, electric, magnetic, and
scalar potentials, respectively.
The variation of (\ref{eq:lag1}) with respect to the potentials leads
to the  Klein--Gordon equation
\be
D^2\kappa + {1\over \rho} D\rho D\kappa -{1\over \kappa} D\kappa^2
-{\alpha^2\over 4f}\left(\kappa^3 D\psi^2 -{1\over \kappa} D\chi^2\right)
=0 \ ,\label{eq:kg}
\ee
the Maxwell equations
\bea
D^2\psi +\left({D\rho \over \rho}  + 2 {D\kappa\over\kappa}
-{Df\over f} \right) D\psi -{1\over \kappa^2 f}(D\epsilon + \psi D\chi)D\chi
&=&0 \ , \\
D^2\chi + \left({D\rho \over \rho}- 2{D\kappa\over \kappa} - {Df\over f}
 \right)D\chi +{\kappa^2\over f} (D\epsilon +\psi D\chi)D\psi &=&0\ ,
\eea
and the Einstein equations
\bea
D^2f + {1\over f}[(D\epsilon + \psi D\chi)^2 - Df^2] + {1\over 2\kappa^2}
(\kappa^4 D\psi^2 + D\chi^2) &=& 0 \ , \\
D^2\epsilon + D \psi D\chi + \psi D^2\chi + (D\epsilon + \psi D\chi)
\left({D\rho\over \rho} -2{D f\over f}\right) &=& 0 \ . \label{eq:ee}
\eea

Notice that the latter system of differential
equations and the Lagrangian (\ref{eq:lag1}) do not contain the function $k$
explicitly.
This is due to the fact that $k$ is a cyclic coordinate of the original
Lagrangian (\ref{lag}) and, as pointed above, a Legendre transformation
has been
used which eliminates all the terms containing the respective ``velocity"
$Dk$ from the Lagrangian (\ref{eq:lag1}). As mentioned above, from the
original field
equations we obtain two differential equations for $k$ that may be written
in terms of the coordinates of the abstract space in the following form
\bea
k_\rho &=&  {\rho \over {4\,f^2}}\,\big[ f_\rho^2-f_z^2 +
\epsilon_\rho^2-\epsilon_z^2 +
({f\over \kappa^2} + \psi^2)\,(\chi_\rho^2-\chi_z^2)\nonumber \\
& & -2\,\psi\,(\epsilon_\rho\,\chi_\rho- \epsilon_z\,\chi_z )+
\kappa^2\,f\,(\psi_\rho^2-\psi_z^2) +
({2\,f\over {\alpha\,\kappa}})^2\,(\kappa_\rho^2-\kappa_z^2)\big]
\label{eq:kar}\\
k_z  &=&{\rho\over 2f^2}\big[
   f_\rho f_z +\epsilon_\rho\epsilon_z +f\psi_\rho\psi_z
-\psi(\epsilon_\rho\chi_z +\epsilon_z\chi_\rho)
+({ f\over \kappa^2 } +\psi^2) \chi_\rho\chi_z
+ ({2f\over \alpha\kappa})^2 \kappa_\rho\kappa_z \big].
 \label{eq:kaz}
\eea

Since Eqs.~(\ref{eq:kg}-\ref{eq:ee})
follow from the
Euler--Lagrange equation applied to the Lagrangian (\ref{eq:lag1}) and the
latter
can be considered as a line element in the potential space, then we conclude
that any solution to the field equations (\ref{eq:lag1}--\ref{eq:kg}) may
be interpreted as a
geodesic in a five--dimensional abstract space with ``coordinates" $f, \
\epsilon, \ \psi,\ \chi$ and $\kappa$. This interpretation is of special
interest since one can use the symmetries of the geodesics generated by
(five--dimensional) Killing vectors, affine collineations, etc. in order
to relate different solutions in the (four--dimensional) spacetime.
However, this task is beyond the scope of this work and will be treated
in further investigations.

\section{Exact Solutions}

It is still very difficult to find exact solutions to the system of coupled
differential equations (\ref{eq:kg}-\ref{eq:ee}). In a recent
work \cite{Ton1}, Matos investigated
a special case ($\alpha =\sqrt{3})$ of the Lagrangian (\ref{lag}) in five
dimensions
and found out that the solution of the main field equations can be reduced to
an equivalent linear problem, based on the Lax pair representation, which
leads to a chiral equation for certain combinations of the spacetime metric
functions or, equivalently, of the potentials in the abstract space. The
respective chiral equation turns out to be invariant with respect to
$SL(3,R)$--transformations and hence solutions of the field equations can be
classified with respect to the subgroups of $SL(3,R)$.  It can be shown
that the existence of a chiral equation is related to the  geometric
properties of the potential space. In fact, the reduction to the problem
to a chiral equation is allowed only if the potential space corresponds
to a Riemannian symmetric space (vanishing of the covariant derivative
of the Riemann curvature tensor).

Unfortunately, it is not possible to adapt the results of \cite{Ton1} in the
present case because the Riemannian space defined by Eq.~(\ref{eq:lag1})
is not symmetric
and, consequently, no chiral equation exists that could be used to
simplify the field equations (\ref{eq:kg}-\ref{eq:ee}) .
It turns out that this property holds only in the special cases $\alpha
= 0, \sqrt{3}$. However, we were able to generalize some of the special
solutions presented in \cite{Ton1} to include the case of arbitrary $\alpha$.
Without
details of calculations we present the resulting solutions and briefly
comment their  properties.

First, we present the general static axisymmetric solution with an arbitrary
harmonic dilaton field. The line element is that of Eq.~(\ref{eq:lel}) and the
potentials are given as
\be
f= e^{\gamma \lambda} \ , \qquad \kappa^2= \kappa_0 e^{\beta\lambda}\ ,
\qquad \epsilon=\psi=\chi=0 \ ,\label{eq:f01}
\ee
where $\kappa_0$ is an arbitrary constant, $\beta$ and $\gamma$ are constants
related by
\be
\beta = {2\alpha^2\over 1+\alpha^2}\ , \qquad \gamma={2\over 1+\alpha^2}  \
 , \label{eq:bet1}
\ee
with $\alpha$ being the arbitrary parameter entering the
Lagrangian (\ref{lag}). Furthermore, $\lambda=
\lambda(\rho,z)$  is a harmonic function, i.e. it satisfies the
two--dimensional Laplace equation
\be \lambda_{,\rho\rho} +{1\over \rho} \lambda_{,\rho} + \lambda_{,zz} = 0 \ .
\ee
According to Eqs.~(\ref{eq:kar}) and (\ref{eq:kaz}), the metric
function $k$ can be calculated for any given $\lambda$ by means of
\be  k_{,\rho} = {\gamma\over 2} \rho (\lambda_{,\rho}^2 - \lambda_{,z}^2)\ ,
\qquad
k_{,z} = \gamma \rho \lambda_{,\rho}\lambda_{,z}\ .
\ee

Equation (\ref{eq:f01}) shows that the harmonic function $\lambda$ determines
the gravitational as well as the dilaton field. In the limiting case $b=0$
($\Phi=0)$, the solution reduces to the well--known Weyl static vacuum
solution \cite{KS}. For an asymptotically flat spacetime, the function
$\lambda$ may be chosen as
\be
\lambda = \sum_{n=0}^\infty q_n
{P_n(\cos\theta)\over (\rho^2 +z^2)^{n+1\over 2} }
\ ,\qquad \cos\theta={z\over (\rho^2+z^2)^{1/2}} \ ,
\ee
where $q_n$ are arbitrary constants and $P_n(\cos\theta)$ are the Legendre
polynomials of order $n$.
A special case of the general solution is that of a
Schwarzschild--like black hole dilaton which corresponds to the choice
$\lambda= \lambda_S= \ln(1-2m/r)$, where $m$=const. and $r$ is a radial
coordinate determined by $\rho = \sqrt{r^2-2mr}\sin\theta $ and
$z = (r-m)\cos\theta$. This solution was first obtained by Janis, Newman
and Winicour \cite{JNW}. They analyzed the behavior of the Schwarzschild
sphere
$r=2m$, showed that it becomes a singular point, and conjectured that the
truncated Schwarzschild solution, where the space suddenly collapses
from a radius slightly greater than $r=2m$ to zero, is a more likely final
state of a generic collapse. This conjecture was further studied by
Christodoulou \cite{Chr}, but as no more independent solutions to the
Einstein--Klein--Gordon field equations are known, the question remains open.
Recently, several time dependent solutions to these field equations has
been found, se for example Ref. \cite{HMN}.
This fact, together with the general solution for an arbitrary
harmonic dilaton presented in this work, will contribute to settle down
the question of the final state of a generic collapse as well as that
about the cosmic censorship conjecture.
For instance, the choice $\lambda= \lambda_S+
\lambda_M$, where $\lambda_M$ is an arbitrary harmonic function may
be used to add a  multipole structure to the Schwarzschild--like
dilaton, a solution that could be used to describe the initial state of
a collapsing configuration.

The generic form of the second class of solutions may be written as
\bea
f = f_0 {e^{\lambda_0\lambda + \tau_0\tau} \over
(a_1\Sigma_1+a_2\Sigma_2)^\gamma }\ ,\quad
&&\kappa^2=\kappa_0^2 (a_1\Sigma_1+a_2\Sigma_2)^\beta
e^{\lambda_0\lambda + (\tau_0-q_1-q_2)\tau}\ , \nonumber\\
\psi&=&{a_3\Sigma_1 + a_4\Sigma_2\over a_1\Sigma_1 + a_2 \Sigma_2} \ ,\quad
\epsilon=\chi=0\ ,  \label{eq:sol1}
\eea
where $a_1, ...,\ f_0,\ q_1, q_2,\ \kappa_0,\ \lambda_0,$ and $\tau_0$ are
constants. Furthermore, $\lambda=\lambda(\rho,z)$ and $\tau=\tau(\rho,z)$ are
harmonic functions; $\Sigma_1$ and $\Sigma_2$ represent functions that can be
given in terms of $\tau$.  In fact, Eq.~(\ref{eq:sol1}) contains two further
subclasses
that follow from the different values of the functions $\Sigma_1$ and
$\Sigma_2$. The first one corresponds to the choice
\be
\Sigma_1 = e^{q_1\tau}\ , \qquad \Sigma_2 = e^{q_2\tau} \ .\label{eq:si1}
\ee
In this case, the constants entering the solution (\ref{eq:sol1}) are
related by
\be
4a_1a_2f_0 +\kappa_0^2 (1+\alpha^2)(a_1a_4-a_2a_3)^2 =0 \ . \label{eq:cons}
\ee
The electric potential $\psi$ is completely determined by the harmonic
function $\tau$ which always appears in the exponential function. To obtain
the second subclass of solution (\ref{eq:sol1}), we must restrict the
values of
the constants $q_1$ and $q_2$, and specify $\Sigma_1$ and $\Sigma_2$ as
follows
\be
q_1=-q_2\ , \qquad {\hbox {and}}\qquad \Sigma_1=\tau\ , \quad
\Sigma_2 = 1\ . \label{eq:si2}
\ee
In contrast to subclass (\ref{eq:cons}), the electric potential $\psi$ is now
given as the ratio of two linear combinations of $\tau$. For the
choice (\ref{eq:si2}), the constants satisfy the relationship
\be
4 a_1^2f_0 - \kappa_0^2 (1+\alpha^2)(a_1a_4-a_2a_3)^2 =0 \ .
\ee

The generic solution (\ref{eq:sol1}) contains two arbitrary harmonic
functions and a large number of constants. This allows us to generate
particular solutions
with totally different properties. The electric potential $\psi$ can be
made to correspond to any desired electric multipole structure (monopole,
dipole, quadrupole, etc.) by
appropriately choosing the value of the harmonic function $\tau$. Once
$\tau$ has been fixed, it still remains the freedom of the function $\lambda$
which can be used to fix the gravitational potential $f$ such that it
describes an arbitrary mass multipole configuration. Consequently, the
harmonic functions $\lambda$ and $\tau$ allow us to ``construct"  any
arbitrary combination of mass and electric multipoles. So far only
electrically charge solutions may be derived from the generic solution
(\ref{eq:sol1}). However, it is easy to see that the field equations
(\ref{fe1}--\ref{fe3}) are invariant under the transformation
\be
\Phi \rightarrow -\Phi\ ,\qquad
F_{\mu\nu} \rightarrow F^*_{\mu\nu} = {1\over 2} e^{-2\alpha\Phi}
\eta_{\mu\nu\rho\sigma} F^{\rho\sigma} \ , \label{eq:trn}
\ee
where $\eta_{\mu\nu\rho\sigma}$ is the Levi--Civita pseudotensor. This
duality rotation, which also involves the dilaton field, may be used
to generate the magnetic counterpart of the generic class (\ref{eq:sol1}).
Accordingly, for any particular electrically charged solution contained in
(\ref{eq:sol1}), the magnetically charged solution may be obtained by
changing the sign of the dilaton.

Now we derive some special solutions contained in (\ref{eq:sol1}).
Consider the
subclass (\ref{eq:si2}) with the following special values of the constants
\be
\tau_0=\lambda_0 = a_3 = 0\ , \qquad
a_1=a_2 = f_0 = \alpha=1=\kappa_0^2 =1 \ .
\ee
According to Eqs.~(\ref{eq:si2}), the resulting solution takes the form
\be
f= {1\over 1+\tau}\ ,\qquad  \kappa^2 = 1+\tau \ , \qquad
\psi = {\sqrt{2}\over 1+\tau} \ ,
\ee
which is equivalent to the static dilatonic version of the Kastor--Trashen
\cite{KT}, \cite{HH2} solution for the value
$\tau = \sum 2M_i/|r-r_i|$. Here $M_i$ and $r_i$ are positive constants,
the sum is over all positive integer values of $i$, and $r$ represents
a radial coordinate. This solution represents  a collection of
extremal electrically charged black holes which are static because the
electric repulsion is balanced by both the gravitational and dilatonic
attraction. Originally, this solution was derived in the low energy limit
of string theory ($\alpha=1)$, but letting in (\ref{eq:si2})
$\alpha$ arbitrary
we easily obtain the generalization to Kaluza--Klein and Einstein--Maxwell
theories: $f=(1+\tau)^{-\gamma}\ , \kappa^2= (1+\tau)^\beta\ ,\
\psi = 2(1+\alpha^2)^{-1/2}(1+\tau)^{-1}$, where $\gamma$ and $\beta$ are
given as in Eq.~(\ref{eq:bet1}). This new exact solution could be
important in
the  understanding of the event horizon, since constant $\alpha$ obviously
adds new structure to the horizon of the static case of the
Kastor--Trashen solution.

Another interesting special solution may be obtained by considering the
subclass (\ref{eq:si1}) with the following values of the constants
\be
f_0=\kappa_0^2 = 1, \quad q_1 = \lambda_0 = 0, \quad \tau_0 = q_2 ,
\quad a_1 = 1- a_2 = {r_+\over r_+ - r_-}\ , \quad a_4 = -a_3
= {2Q\over r_+ - r_-}\ ,
\ee
where $r_+$, $r_-$, and $Q$ are constants. With this choice of the constants
the harmonic function $\lambda$ does not enter the resulting solution. We
fix the remaining function $\tau$ as
\be
\tau = {1\over q_2} \ln \left(1 + {r_--r_+\over r}\right) \ ,
\ee
where $r$ is a radial coordinate related to cylindrical coordinates by
means of $\rho = \break \sqrt{r^2 +(r_--r_+)r}\sin\theta$ and
$z=[r+(r_--r_+)/2]\cos\theta$. Inserting these values into
Eqs.~(\ref{eq:sol1}) and \ref{eq:si1}, we obtain
\be
f= \left(1 + {r_--r_+\over r}\right)\left(1+{r_-\over r}\right)^{-\gamma}
\ ,\quad \kappa^2 = \left(1 + {r_-\over r}\right)^\beta \ ,
\psi = -{2Q\over r+r_-}\ .
\ee
This solution was originally obtained by Gibbons and Maeda \cite{GM}; the
representation given here coincides with that of Horne and
Horowitz \cite{HH1}
(after performing the coordinate displacement $r\rightarrow r+r_-$ in
Eqs.~(6--10) of Ref.\cite{HH1}).

Another special solution contained in (\ref{eq:sol1}) was recently given by
Matos \cite{Ton2} for an arbitrary magnetic field coupled to the dilaton. It
is direct to see that it is contained in (\ref{eq:sol1}) by using the
dual transformation
given in Eq.~(\ref{eq:trn}), and setting $\tau_0=0$ and $\alpha=\sqrt{3}$.

All the special solutions given here contain other particular solutions in
each of theories that can be obtained by specifying the value of $\alpha$.
We believe that all known static solutions of the field equations
following from the Lagrangian (\ref{lag}) are contained in
Eqs.~(\ref{eq:sol1}), (\ref{eq:si1}) and (\ref{eq:si2}) as special cases.
It would be interesting to find stationary solutions
with both electric and magnetic field. This could be done by using the
symmetry properties of the metric (\ref{eq:lag1}) of the potential as
mentioned above.
In fact, we obtained all the Killing vectors of metric (\ref{eq:lag1}) but,
unfortunately,
they cannot be used to generate solutions since they correspond to ``gauge
transformations" of the potential, i.e., the ``stationary electromagnetic"
solutions generated by applying infinitesimal Killing transformations
on the solution (\ref{eq:sol1}) contain only trivial angular momentum and
magnetic fiel terms that can be eliminated by appropriate coordinate
transformations. Consequently, we have to derive  more general symmetry
properties of the potential space (affine collineations, curvature
collineations, etc.). This task will be treated in further investigations.

\end{document}